# Local heat emission due to unidirectional spin-wave heat conveyer effect observed by lock-in thermography

Yuta Kainuma[1,a)], Ryo Iguchi[2,a,b)], Dwi Prananto[1], Vitaliy I. Vasyuchka[3], Burkard Hillebrands[3], Toshu An[1,b)], and Ken-ichi Uchida[2,4,5]

[1]*School of Materials Science, Japan Advanced Institute of Science and Technology, Nomi, Ishikawa 923-1292, Japan*
[2]*National Institute for Materials Science, Tsukuba 305-0047, Japan*
[3]*Fachbereich Physik and Landesforschungszentrum OPTIMAS, Technische Universität Kaiserslautern, 67663 Kaiserslautern, Germany*
[4]*Institute for Materials Research, Tohoku University, Sendai 980-8577, Japan*
[5]*Center for Spintronics Research Network, Tohoku University, Sendai 980-8577, Japan*

a) Y. Kainuma and R. Iguchi contributed equally to this work.
b) Author to whom correspondence should be addressed: IGUCHI.Ryo@nims.go.jp and toshuan@jaist.ac.jp

**Abstract:**
Lock-in thermography measurements were performed to reveal heat source distribution induced by the unidirectional spin-wave heat conveyer effect (USHCE) of magnetostatic surface spin waves. When the magnetostatic surface spin waves are excited in an yttrium iron garnet slab, the lock-in thermography images show spatially biased sharp and complicated heating patterns, indicating the importance of edge spin-wave dynamics for USHCE. The accessibility to the local heat emission properties allows us to clarify a capability of remote heating realized by USHCE; it can transfer energy for heating even through a macro-scale air gap between two magnetic materials owing to the long-range dipole-dipole coupling.

During the last decade, the role of spins in thermal and thermoelectric transport phenomena has been investigated in the field of spin caloritronics.[1–3] In particular, spin waves, collective dynamics of spins,[4–8] are found to be a key energy carrier in spin-mediated thermoelectric conversions, i.e., the spin Seebeck and Peltier effects.[9–14]

Among spin waves, the magnetostatic surface spin wave (MSSW) has unique features: surface localization and non-reciprocal propagation.[4,15] MSSWs exist only near surfaces of magnetic materials and propagate only in one direction specified by the outer product of the magnetization vector and surface normal. MSSWs carry not only spin angular momenta but also energy,[16–19] resulting in heating via dissipative couplings to phonons. Using these features, in 2013, An *et al.* demonstrated that MSSWs can be used for controlled remote heating; this phenomenon is called the unidirectional spin-wave heat conveyer effect (USHCE) [Fig. 1(a)].[20]

USHCE has been experimentally investigated in ferrimagnetic yttrium iron garnet ($Y_3Fe_5O_{12}$: YIG) slabs by means of infrared thermography, where the nonequilibrium steady-state temperature distribution is captured through thermal radiation with an infrared camera. The steady-state thermal images clearly show that the MSSW excitation results in heating far from the excitation source and the heating area can be changed by the magnetic field direction [Fig. 1(a)].[20] However, the position and distribution of USHCE-induced heating sources are remained to be clarified. Recently, an extended thermographic technique called the lock-in thermography (LIT) has been introduced to the study of spin caloritronics.[12,21–26] Owing to the high temperature resolution of LIT (< 1 mK), USHCE was observed also in thin YIG films.[22,23] In addition to the improved sensitivity, LIT enables transient thermal imaging measurements at frequencies over 10 Hz, which is high enough to suppress spatial broadening of temperature change due to thermal conduction for millimeter-scale systems.[26] Thus, LIT allows us to reveal the distribution of the heat sources due to USHCE.

In this study, we performed LIT measurements of USHCE in a millimeter-scale YIG slab. Heating due to MSSWs is repeated periodically in time and the distribution of the resultant temperature change is imaged [Fig. 1(b)]. By increasing the frequency of the temperature modulation, we reveal the local heat emission from MSSWs. We also investigated the effect of the long-range interacting nature of MSSWs on USHCE by performing the same experiment using two YIG slabs separated by an air gap.

The sample used in this study is a polycrystalline sintered YIG slab with a rectangular of $13.6 \times 1.9 \times 1.0$ mm$^3$. Spin waves were excited using a 25-μm-thick wire antenna located at the center of the top surface of the sample. The antenna wire was soldered to the signal lines of two 3-mm-wide microstrip lines, of which the characteristic impedance is 50 Ω. A magnetic field with its magnitude $H$ was applied along the microwave antenna to form the MSSW geometry.[4,5,16–18] For spin-wave spectrum acquisition, the microstrip lines were connected to a vector network analyzer and the reflection spectrum $S_{11}$ was measured at the microwave power $P = 1$ mW. For the LIT measurement, one side of the microstrip lines was connected to a signal generator via a microwave amplifier and the other side to a 50 Ω load, where $P$ was chopped at the lock-in frequency $f_L$ [Fig. 1(b)]. In LIT, thermal images of 640 × 480 pixels, which approximately corresponds to a viewing area of $20.4 \times 15.3$ mm$^2$, were captured at a framerate of 60 Hz by an infrared camera. Then, the sinusoidal wave component of temperature modulation oscillating at $f_L$ was extracted through the Fourier analysis, producing the amplitude $A$ and phase delay $\varphi$ (due to thermal conduction) images. In the $A$ and $\varphi$ images, the temperature modulation synchronized with the chopped microwave power is visualized [Fig. 1(b)].[21] Differently from the previous studies, the sample surface was not covered by a black ink because YIG has high emissivity in the wavelength range of our infrared detector (7.5 to 14.0 μm).[25] In this setup, we can avoid propagation loss of spin waves due to a black ink. In the following, we show the results measured at $\mu_0 H$ = 95.0 mT and $P$ = 160 mW, where $\mu_0$ is the vacuum permeability. We confirmed the linearity of the spin-wave excitation up to $P$ = 160 mW by microwave spectroscopy measurements and that of USHCE by the LIT measurements.

The excitation of spin waves was confirmed by the change in the reflection spectra, $\Delta|S_{11}|^2$, which is the difference between the $|S_{11}|^2$ values with and without the magnetic field. The spin-wave modes can be speculated from the shape of the $\Delta|S_{11}|^2$ spectrum;[20] the observed spectrum is composed of peaks due to the magnetostatic backward volume waves (MSBVWs), the uniform precession mode [i.e., ferromagnetic resonance (FMR)], and MSSWs. The FMR mode appears at the lowest frequency edge of the band of MSSWs and does not propagate.[5] At the frequencies lower than the FMR condition, MSBVW can be excited when standing waves are formed and coupled to the antenna's form factor.[20,27] As shown in Figs. 2(b) and 2(c), the steady-state thermal images show the existence of USHCE only when MSSWs are excited.

Figures 2(b) and 2(c) show the $f_L$ dependence of the USHCE-induced temperature modulation for the modes B and E labeled in Fig. 2(a). Under a nearly steady-state condition ($f_L$ = 0.1 Hz), the mode B (E) shows the gradual temperature modulation symmetric (asymmetric) about the excitation antenna; these temperature changes are in good agreement with that for FMR (MSSW) observed in the previous study.[20] Importantly, with increasing $f_L$, the temperature distribution becomes sharp or focused, revealing the local heat emission by suppressing temperature broadening due to thermal conduction.[21] The heat sources for the mode B were found to be localized near the antenna and the side edges of the sample, whereas, interestingly, those for the mode E (a peak in the MSSW band) locate in the center lane and reach the area close to the sample end. These results demonstrate that USHCE induces heat sources even far from the excitation antenna.

Figure 3 shows the thermal images for various spin-wave modes [labels A–H in Fig. 2(a)] measured at the steady state and high lock-in frequency ($f_L$ = 15.0 Hz). The steady-state images show that, as the microwave frequency $f_{mw}$ increases, the temperature-modulation distributions become asymmetric about the antenna, being consistent with the spectral features: reciprocal MSBVW and FMR at low $f_{mw}$ and non-reciprocal MSSW at high $f_{mw}$. The images at high $f_L$ values show two features. First, the localized heat sources exist near the side edges. This is recognized for most of the modes and prominent for the modes A–C, F, and G, indicating that this feature is not exclusive for MSSWs. Second, the MSSW modes show the beam- and circular-shaped heat sources on the center lane. Starting from the mode D, the beam-shaped source elongates. At the mode F, the length significantly decreases, while an isolated circular-shaped heat source appears near the sample end.

The revealed complex heat-source distribution shows the importance of spin-wave dynamics in USHCE. Here, we discuss the possible origin of the two main features in the results: the side-edge heat sources and the various heating patterns due to MSSWs. The former feature is attributed to the direct and/or indirect excitation of edge-mode spin waves. At sample edges, the effective magnetic potential is reduced, so that the spin waves can be confined.[28–30] Such edge modes are expected to be immobile, and thus they can act as a trap for the excited spin waves; once the excited spin-wave mode is converted into an edge mode, it cannot travel anymore and eventually deposits its energy at its position on a time scale given by its relaxation time. The latter feature, the variation of heating patterns due to MSSWs, cannot be discussed by a simple scenario. Since the propagation length of MSSW depends on the group velocity and/or relaxation time, the length change of the bar-shaped heat source for the modes C–E in Fig. 3 is reasonably expected. On the other hand, the appearance of the isolated heat sources for the modes F–G cannot be explained by this scenario. Thus, one needs to consider the dynamics of MSSWs at the sample end. Because MSSWs propagating in the opposite directions have a low profile overlap because of their nonreciprocal nature,[15,31] the backscattering of MSSWs at the sample end is suppressed.[20] Consequently, MSSWs excited on the top surface are first converted into standing spin-wave modes and then into MSSWs on the bottom surface.[32] This process takes time and thus results in focusing of the energy of spin waves at the sample end. We also note that the focusing of MSSWs can occur when the effective magnetic potential is gradually changed near the end.[33] To theoretically describe USHCE, such spin-wave dynamics needs to be incorporated into the heat conduction calculation, for example, as reported in Ref. 34. To experimentally verify these scenarios, the combination between LIT and other techniques for observing spin-wave dynamics[8,18,35–37] is useful.

As LIT allows us to determine the heat source distribution, we can further verify the remote heating properties realized by USHCE. As an example of the features unique to USHCE, we demonstrate that USHCE can generate heat even when an air gap is introduced in a medium owing to the long-range dipole-dipole interaction of MSSWs.[5,38,39] Here, we measured the temperature modulation induced by USHCE in two rectangular YIG slabs separated by the distance $d$ [see Fig. 4(a)]. The dimensions of each slab are $4.5 \times 1.9 \times 1.0$ mm$^3$ and the microwave wire antenna is located on one end of the lower YIG slab [solid white line in Fig. 4(b)]. Spin waves were excited at $P = 160$ mW and $\mu_0 H = 95$ mT, where $f_{mw}$ was chosen for the heating in the upper slab to be recognized and the temperature modulation profiles for various $d$ values to be similar to each other. As shown in Fig. 4(b), for $d = 0$ mm, we observed USHCE signals in both steady-state and LIT thermal images. Interestingly, a clear isolated circular-shaped heat source was observed to appear even in the presence of the air gap [see the enlarged $A$ images shown in Fig. 4(c)]. The $\varphi$ values for the remote temperature modulation are the same as those around the excitation antenna. Thus, these signals originate not from the thermal conduction but from the MSSW propagation. As shown in Fig. 4(d), the averaged amplitude $A_{avg}$ value in the white rectangles in Fig. 4(c) maintains its magnitude up to $d \sim 1.0$ mm. Appearance of the peak at $d = 0.5$ mm may be attributed to the competition of the transmittance over the air gap and the excitation efficiency at the antenna [see the $T$ and $A$ images in the lower slab in Fig. 4(b)], although further experimental improvement on the signal to noise ratio is necessary for the quantitative analysis. Nevertheless, these results confirm that USHCE can be used even when air gaps are introduced, which is a unique feature brought by the long-range nature of MSSWs.

In summary, local heat emission due to USHCE has been investigated in YIG slabs by means of LIT.

High-frequency LIT measurements allow us to resolve the distribution of the USHCE-induced heat sources. We found the two interesting features in the USHCE-induced temperature modulation: the localized heat sources around the side edges of the slab and isolated heat sources near the end of the slab. In addition, USHCE over an air gap has been demonstrated. We anticipate that the revealed characteristics of USHCE will be useful for constructing unconventional thermal management technologies based on spin waves.

**Acknowledgments**

This work was supported by CREST "Creation of Innovative Core Technologies for Nano-enabled Thermal Management" (JPMJCR17I1) from JST, Japan, and Joint Research Hub Program from NIMS, Japan.

**DATA AVAILABILITY**
The data that support the findings of this study are available from the corresponding authors upon reasonable request.

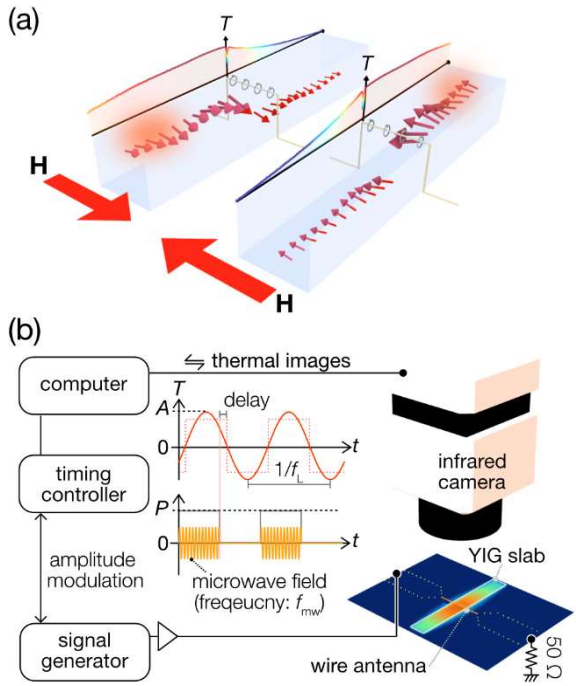

**Fig. 1.**

(a) Schematic of USHCE due to MSSWs excited by a microwave antenna. The localization surface and propagation direction of MSSWs are changed by the reversal of the magnetic field (**H**) direction, which results in different temperature ($T$) profiles. (b) Measurement setup for LIT consisting of a computer, an infrared camera, a signal generator, and a microwave amplifier. For observing local heat emission due to spin waves excited in a YIG slab, the microwave power $P$ was pulsed with the lock-in frequency $f_L$. The resultant lock-in amplitude $A$ and phase delay $\varphi$ of the first harmonic sinusoidal component (the red line in the $T$ plot) of the spin-wave-induced temperature modulation were imaged. The infrared camera captures the temperature distribution on a frame of $13.6 \times 1.9$ mm$^2$ on the surface of the YIG slab.

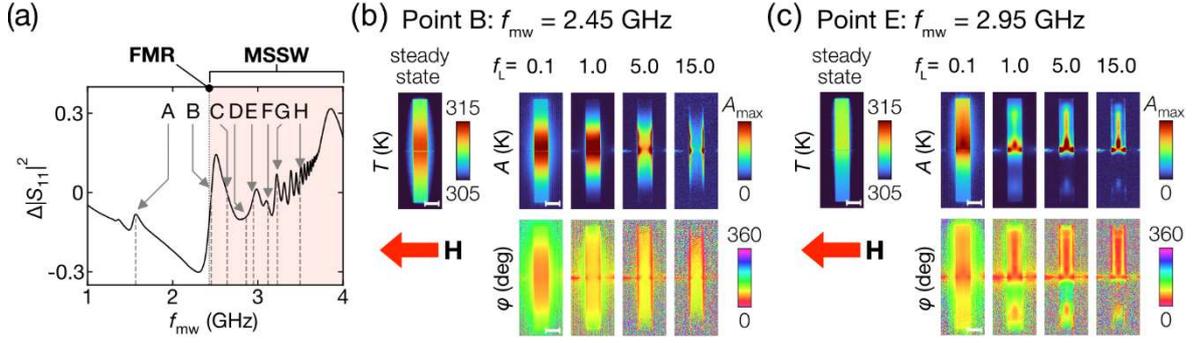

**Fig. 2.**

(a) Microwave reflection spectrum change ($\Delta|S_{11}|^2$) obtained using a vector network analyzer at $P$ = 1 mW and $\mu_0 H$ = 95 mT. The spin-wave band of MSSWs is colored to be pink. The frequencies chosen for the LIT measurements were labelled as A–H. (b, c) Steady-state temperature $T$ images and $f_L$ dependence of the lock-in amplitude $A$ and phase $\varphi$ images for (b) the FMR mode (B: $f_{mw}$ = 2.45 GHz) and (c) the MSSW mode (E: $f_{mw}$ = 2.95 GHz). For the FMR mode (b), the maximum limit in the amplitude plot, $A_{max}$, is 1.60, 0.20, 0.06, and 0.03 K for $f_L$ = 0.1, 1.0, 5.0, and 15.0 Hz, respectively. For the MSSW mode (c), $A_{max}$ is 1.00, 0.30, 0.10, and 0.04 K for $f_L$ = 0.1, 1, 5, and 15.0 Hz, respectively. The length of the scale bar (white) is 2 mm.

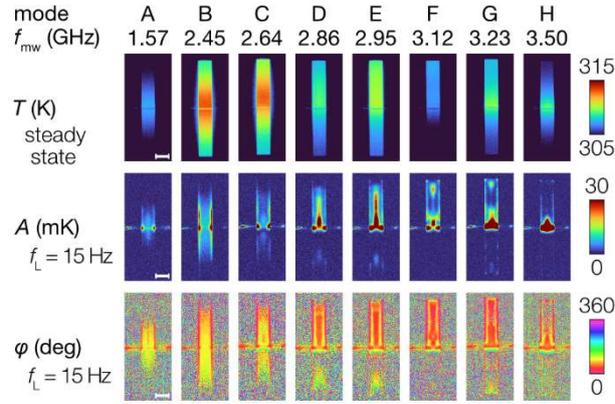

**Fig. 3.**

Steady-state and LIT images for various values of $f_{mw}$ at $f_L$ = 15.0 Hz. The labels A – H correspond to the positions denoted in the $\Delta|S_{11}|^2$ spectrum in Fig. 2(a). The length of the scale bar (white) is 2 mm.

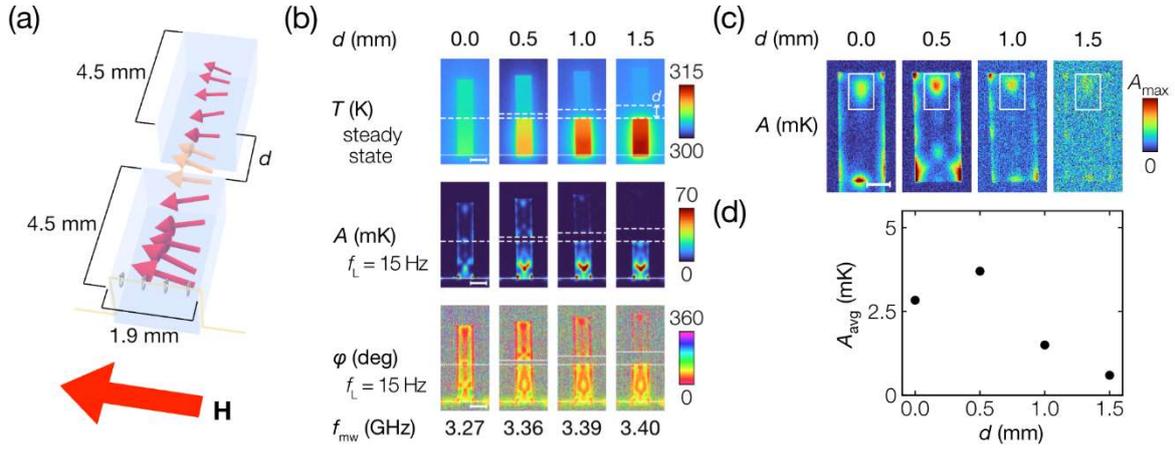

**Fig. 4**.

(a) Schematic of the experimental configuration used for demonstrating the USHCE-induced heating through an air gap. $d$ denotes the distance between the two rectangular YIG slabs. MSSWs were excited at one end of the lower YIG slab using a microwave wire antenna. (b) Steady-state and LIT images at $f_L$ = 15.0 Hz for $d$ = 0, 0.5, 1.0, and 1.5 mm. The white dotted lines (solid lines) represent the position of the air gaps (the antenna). (c) Magnified views of the $A$ images for the upper YIG slab. Here, the maximum limit of the plot, $A_{max}$, is 20, 20, 10, and 5 mK for $d$ = 0, 0.5, 1.0, and 1.5 mm, respectively. (d) $d$ dependence of the amplitude $A_{avg}$ averaged over the white rectangles in (c). The length of the scale bar (white) is 2 mm for (b) and 1 mm for (c).